\documentstyle[11pt,epsfig]{article}
\date{}
\begin{document}
\title{{\bf Bianchi type I cyclic cosmology from Lie-algebraically deformed phase space}}
\author{Babak Vakili$^{1}$\thanks{email: b-vakili@iauc.ac.ir}\,\, and\,\,\,Nima Khosravi$^{2}$\thanks{email: n-khosravi@sbu.ac.ir}\,\,\,\\\\$^1${\small
{\it Department of Physics, Azad University of Chalous, P. O. Box
46615-397, Chalous, Iran }}\\$^2${\small {\it Department of
Physics, Shahid Beheshti University, G. C., Evin, Tehran 19839,
Iran}} $^{}${}} \maketitle

\begin{abstract}
We study the effects of noncommutativity, in the form of a
Lie-algebraically deformed Poisson commutation relations, on the
evolution of a Bianchi type I cosmological model with a positive
cosmological constant. The phase space variables turn out to
correspond to the scale factors of this model in $x$, $y$ and $z$
directions. According to the conditions that the structure
constants (deformation parameters) should satisfy, we argue that
there are two types of noncommutative phase space with
Lie-algebraic structure. The exact classical solutions in
commutative and type I noncommutative cases are presented. In the
framework of this type of deformed phase space, we investigate the
possibility of building a Bianchi I model with cyclic scale
factors in which the size of the Universe in each direction
experiences an endless sequence of contractions and reexpansions.
We also obtain some approximate solutions for the type II
noncommutative structure by numerical methods and show that the
cyclic behavior is repeated as well. These results are compared
with the standard commutative case, and similarities and
differences of these solutions are discussed.\vspace{5mm}\newline
 \vspace{0.8mm}\newline
\vspace{.5cm}
\end{abstract}
\section{Introduction}
Since 1947 when noncommutativity between space-time coordinates
was first introduced by Snyder \cite{1}, many efforts have been
made in this area and the corresponding results have been followed
by a number of works, the main motivations of which lie in the
results of string theory \cite{2}. Although the interest in
noncommutative theories has been gathering pace in recent years
because of strong motivations in the development of string and
M-theories, they may also be justified in their own right because
of the interesting predictions they have made in particle physics
\cite{3}, quantum \cite {4} and classical mechanics \cite{5}. In a
general classification there are three sorts of noncommutative
space-time with the following deformed algebra between coordinates
\cite{6}

 $\bullet$ Deformed Poisson algebra with canonical
structure

\begin{equation}\label{MO1}
\{x_{\mu},x_{\nu}\}=\theta_{\mu \nu},\end{equation}where the
tensor (with constant components) $\theta_{\mu \nu}$ is assumed to
be antisymmetric.

$\bullet$ Deformed Poisson algebra with Lie-algebraic structure
\begin{equation}\label{MO2}
\{x_{\mu},x_{\nu}\}=\theta_{\mu
\nu}^{\lambda}x_{\lambda},\end{equation}where the (constant)
deformation parameters $\theta_{\mu \nu}^{\lambda}$ are assumed to
be antisymmetric to lower indices.

$\bullet$ Deformed Poisson algebra with quantum space structure
\begin{equation}\label{MO3}
\{x_{\mu},x_{\nu}\}=\theta_{\mu \nu}^{\alpha \beta}x_{\alpha}x_
{\beta},\end{equation}where again the (constant) deformation
parameters $\theta_{\mu \nu}^{\alpha \beta}$ are assumed to be
antisymmetric to lower indices.

In general, as is clear from (\ref{MO1})-(\ref{MO3}),
noncommutativity in their original form imply a noncommutative
underlying geometry for space-time. However, formulation of
gravity in a noncommutative space-time is highly nonlinear,
rendering the setting up of cosmological models difficult. In this
respect, a different approach to noncommutativity is through its
introduction in the phase space constructed by minisuperspace
fields and their conjugate momenta. Alternatively, in cosmological
systems, since the scale factors, matter fields and their
conjugate momenta play the role of dynamical variables of the
system, introducing noncommutativity in the corresponding phase
space is particularly relevant \cite{7}. This means that in the
relations (\ref{MO1})-(\ref{MO3}) the space-time coordinates can
be replaced by the phase space coordinates to construct a
noncommutative or deformed phase space. In this sense, one can
introduce different kinds of noncommutativity between different
dynamical variables of the corresponding phase space and the main
aim is to study the aspects related to the application of
noncommutativity in the framework of minisuperspace reduction of
dynamics.

In this paper we are going to investigate the impact of the
deformation of Lie-algebraic type on the cosmological dynamics of
Bianchi type I model. Since the Bianchi models have different
scale factors in different directions, they are suitable
candidates for studying noncommutative cosmology. Here, our aim is
to introduce Lie-algebraic noncommutative scale factors in Bianchi
type I space-time and compare and contrast their solutions to that
of the conventional case. It should be emphasized that when we
speak of noncommutativity in this work, we mean noncommutativity
in the fields (scale factors) and not the coordinates, that is to
say that we study noncommutativity within the context of phase
space only. We shall see that the main feature of the resulting
noncommutative model is constructing a scenario of cyclic
cosmology governed by introducting a Lie-algebraic deformation in
the phase space of Bianchi type I space-time. In such a scenario
the history of the Universe (in each direction) is periodic, i.e.,
the Universe undergoes a periodic sequence of expansion and
contraction phases. The existence of cyclic solutions in the FRW
models was first introduced by Tolman in \cite{8} and then the
early results in this subject have been followed by a number of
works, see \cite{9} and for a review see \cite{10}. In these works
cyclic cosmologies which are an extension of bouncing scenarios,
are used as an alternative for conventional inflationary
cosmology. Instead of a continuously expansion after big-bang, the
bang is replaced by a transition to an earlier phase of evolution.

The structure of the paper is as follows: after a brief review of
Bianchi type I cosmology with a cosmological constant in section
2, we deal with a Lie-algebraically deformed phase space in
section 3. In this section by imposing some constraints (Jacobi
identity) on the structure constants of the Lie algebra, we
present two kinds of Lie-algebraic deformed of Poisson brackets
according to which we write the (Poisson) commutation relations
between the dynamical variables of the Bianchi I model. The exact
solutions in type I noncommutative cases are presented. We also
obtain, in this section, some approximate solutions for the type
II noncommutative structure by numerical methods. Finally, in
section 4 we summarize and discuss the results.

\section{Bianchi type I cosmology with cosmological constant}
In this section we make a quick review of some of the important
results in the Bianchi type I model and obtain its Lagrangian and
Hamiltonian in the ADM decomposition, for more details see
\cite{Ryan}. The simplest generalization of the flat FRW cosmology
is the Bianchi type I Universe which has the metric
\begin{equation}\label{A1}
ds^2=-N^2(t)dt^2+a^2(t)dx^2+b^2(t)dy^2+c^2(t)dz^2.\end{equation}In
this metric $N(t)$ is the lapse function and there are three
functions $a(t)$, $b(t)$ and $c(t)$, to be determined by the
Einstein field equations, and are the scale factors of the
corresponding Universe in the $x$, $y$ and $z$ directions
respectively. The scale factors in different directions are
allowed to vary independently of each other. This metric is the
simplest anisotropic and homogeneous cosmological model which,
upon making the scale factors equal, becomes the flat FRW metric.
In general, the nine Bianchi (class A) models are the most general
homogeneous cosmological solutions of the Einstein field equations
which admit a three-dimensional isometry group, i.e. their
spatially homogeneous sections are invariant under the action of a
three-dimensional Lie group. To transform the Lagrangian of the
dynamical system which corresponds to the Bianchi cosmologies to a
more manageable form, it is useful to introduce the following
change of variables
\begin{equation}\label{A2}
a(t)=e^{u(t)+v(t)+\sqrt{3}w(t)},\hspace{0.5cm}b(t)=e^{u(t)+v(t)-\sqrt{3}w(t)},\hspace{0.5cm}c(t)=e^{u(t)-2v(t)}.\end{equation}
In the Misner notation \cite{Misner}, the metric of the Bianchi
models can be written in terms of these new variables as
\begin{equation}\label{A}
ds^2=-N^2(t)dt^2+e^{2u(t)}e^{2\beta_{ij}(t)}\omega^i\otimes \omega^j,
\end{equation}where $V(t)=e^{3u(t)}=abc$ is the comoving volume of the Universe and $\beta_{ij}$ determines the
anisotropic parameters $v(t)$ and $w(t)$ as follows
\begin{equation}\label{B}
\beta_{ij}=\mbox{diag}\left(v+\sqrt{3}w,v-\sqrt{3}w,-2v\right).
\end{equation}Also, in metric (\ref{A}), the one-forms $\omega^i$ represent the invariant one-forms of the corresponding
isometry group and satisfy the following Lie algebra
\begin{equation}\label{C}
d\omega^i=-\frac{1}{2}C^i_{jk}\omega^j
\wedge\omega^k,\end{equation} where $C^i_{jk}$ are the structure
constants. Indeed, the Bianchi models are grouped by their
structure constants into classes A and B. Because of the
difficultly in formulating the class B Bianchi models in the
context of the ADM decomposition and canonical quantization
\cite{Haw}, it is usually the case that one confines attention to
the class A models where the structure constants obey the relation
$C^i_{ji}=0$.

\begin{figure}
\begin{tabular}{c}\hspace{-1cm} \epsfig{figure=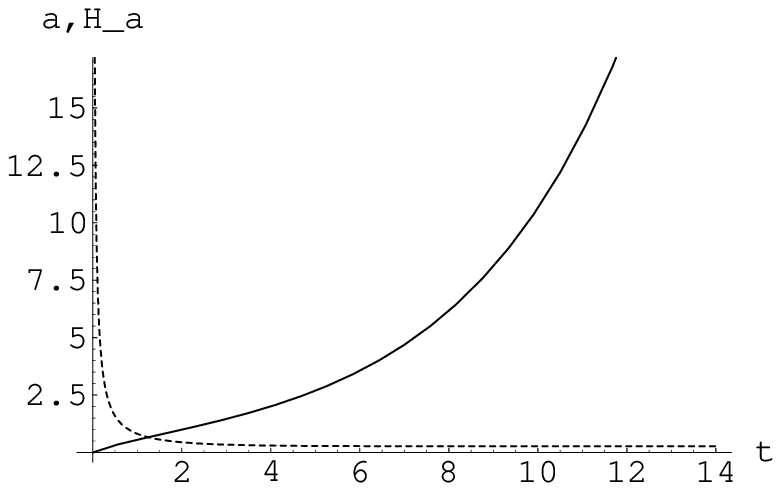,width=5cm}
\hspace{1.5cm} \epsfig{figure=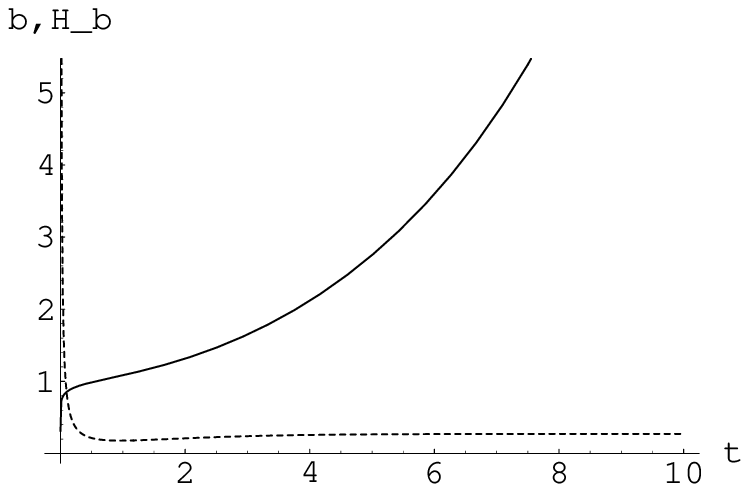,width=5cm}\hspace{1.5cm}
\epsfig{figure=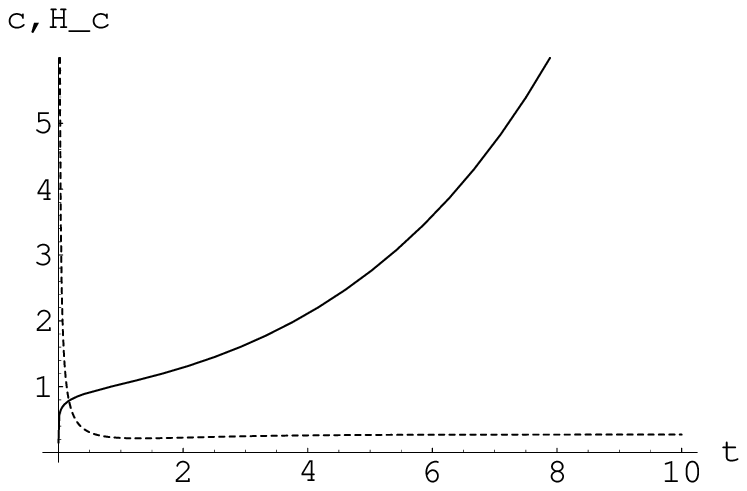,width=5cm}
\end{tabular}
\begin{tabular}{c}\hspace{-1cm} \epsfig{figure=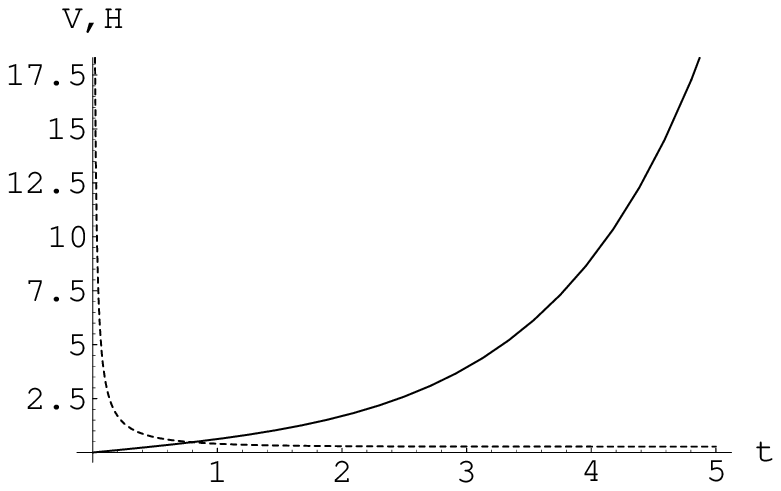,width=5cm}
\hspace{1.5cm} \epsfig{figure=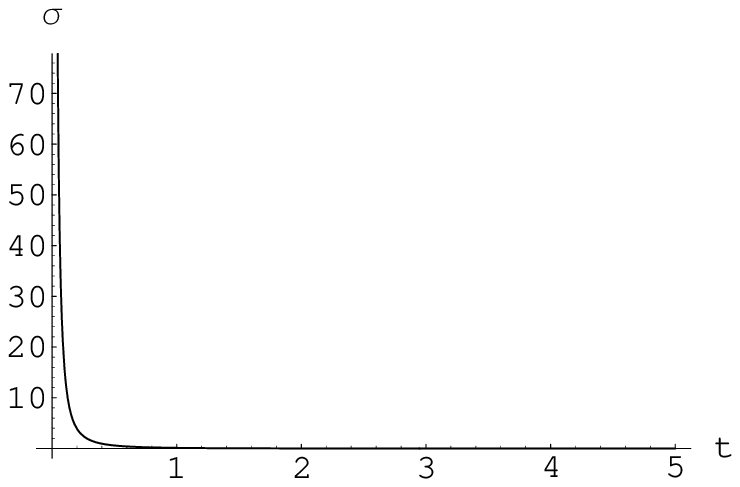,width=5cm}\hspace{1.5cm}
\epsfig{figure=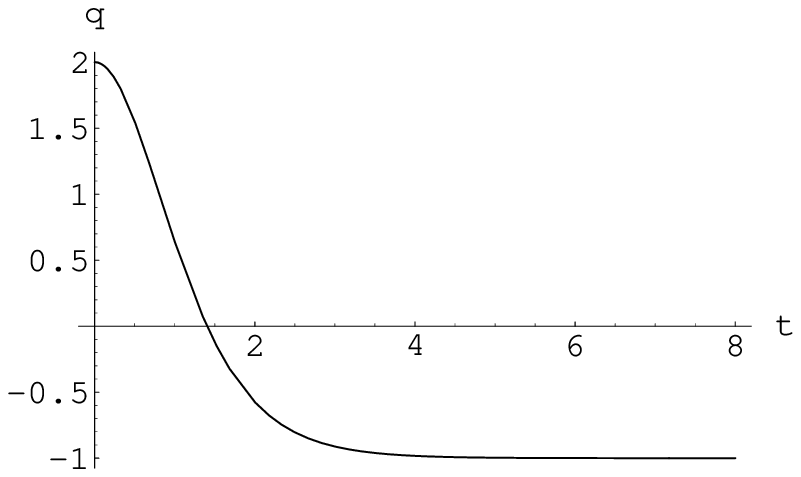,width=5cm}
\end{tabular}
\caption{\footnotesize Top: Behavior of the scale factors $a(t)$,
$b(t)$ and $c(t)$ (solid lines) and their corresponding Hubble
parameters (dashed lines) as a function of time. Bottom: The
comoving volume $V(t)=e^{3u}=abc$ (solid line) and its
corresponding Hubble parameter (dashed line), the shear scalar and
the deceleration parameter. The figures are plotted for numerical
values $\Lambda=1$, $p_{0v}=1$ and $p_{0w}=2$.}\label{fig1}
\end{figure}

The Einstein-Hilbert action is given by (we work in units where
$c=\hbar=16\pi G=1$)
\begin{equation}\label{D}
{\cal S}=\int d^4 x\sqrt{-g}({\cal R}-\Lambda),\end{equation}
where $g$ is the determinant of the metric, ${\cal R}$ is the
scalar curvature of the space-time metric (\ref{A}) and $\Lambda$
is the cosmological constant. In terms of the ADM variables,
action (\ref{D}) can be written as \cite{Kief}
\begin{equation}\label{E}
{\cal S}=\int dt d^3x{\cal L}=\int dt d^3x
N\sqrt{h}\left(K_{ij}K^{ij}-K^2+R-\Lambda\right),\end{equation}
where $K_{ij}$ are the components of extrinsic curvature (second
fundamental form) which represent how much the spatial space
$h_{ij}$ is curved in the way it sits in the space-time manifold.
Also, $h$ and $R$ are the determinant and scalar curvature of the
spatial geometry $h_{ij}$ respectively, and $K$ represents the
trace of $K_{ij}$. The extrinsic curvature is given by
\begin{equation}\label{F}
K_{ij}=\frac{1}{2N}\left(N_{i|j}+N_{j|i}-\frac{\partial
h_{ij}}{\partial t}\right),\end{equation}where $N_{i|j}$
represents the covariant derivative with respect to $h_{ij}$.
Using (\ref{A}) and (\ref{B}) we obtain the nonvanishing
components of the extrinsic curvature and its trace as follows
\begin{eqnarray}\label{G}
\left\{
\begin{array}{ll}
K_{11}=-\frac{1}{N}(\dot{u}+\dot{v}+\sqrt{3}\dot{w})e^{2(u+v+\sqrt{3}w)},\\\\
K_{22}=-\frac{1}{N}(\dot{u}+\dot{v}-\sqrt{3}\dot{w})e^{2(u+v-\sqrt{3}w)},\\\\
K_{33}= -\frac{1}{N}(\dot{u}-2\dot{v})e^{2(u-2v)},\\\\
K=-3\frac{\dot{u}}{N},
\end{array}
\right.
\end{eqnarray}
where a dot represents differentiation with respect to $t$. The
scalar curvature $R$ of a  spatial hypersurface is a function of
$v$ and $w$ and can be write in terms of the structure constants
as \cite{Chris}
\begin{equation}\label{H}
R=C^i_{jk}C^l_{mn}h_{il}h^{km}h^{jn}+2C^i_{jk}C^k_{li}h^{jl}.
\end{equation}
The Lagrangian for the Bianchi class A models may now be written by substituting the
above results into action (\ref{E}), giving
\begin{equation}\label{I}
{\cal
L}=\frac{6e^{3u}}{N}\left(-\dot{u}^2+\dot{v}^2+\dot{w}^2\right)+Ne^{3u}\left(R-\Lambda
\right).
\end{equation}
The momenta conjugate to the dynamical variables are given by
\begin{equation}\label{J}
p_u=\frac{\partial {\cal L}}{\partial
\dot{u}}=-\frac{12}{N}\dot{u}e^{3u},\hspace{.5cm}p_v=\frac{\partial
{\cal L}}{\partial
\dot{v}}=\frac{12}{N}\dot{v}e^{3u},\hspace{.5cm}p_w=\frac{\partial
{\cal L}}{\partial \dot{w}}=\frac{12}{N}\dot{w}e^{3u},
\end{equation}
leading to the following Hamiltonian
\begin{equation}\label{K}
{\cal
H}=\frac{1}{24}Ne^{-3u}\left(-p_u^2+p_v^2+p_w^2\right)-Ne^{3u}\left(R-\Lambda\right).
\end{equation}The preliminary set-up for writing the dynamical equations is now complete.
The cosmological dynamics of the Bianchi models are studied in
many works  \cite{Ryan}-\cite{Her}. In the following, we shall
consider only the simplest Bianchi class A model, namely type I.
The structure constants of the Bianchi type I are all zero, that
is $C^i_{jk}=0$. It then follows from equation (\ref{H}) that $R =
0$. Thus, with the choice of the cosmic time gauge $N =1$, the
Hamiltonian can be written as
\begin{equation}\label{L}
{\cal H}=\frac{1}{24}e^{-3u}\left(-p_u^2+p_v^2+p_w^2\right)+\Lambda
e^{3u}.
\end{equation}The Poisson brackets for the phase space variables are
\begin{equation}\label{M}
\{x_i,x_j\}=0,\hspace{.5cm}\{p_i,p_j\}=0,\hspace{.5cm}\{x_i,p_j\}=\delta_{ij},
\end{equation}
where $x_i(i=1,2,3)=v,w,u$ and $p_i(i=1,2,3)=p_v,p_w,p_u$. Therefore, the classical dynamics is governed by the Hamiltonian equations, that is
\begin{eqnarray}\label{O}
\left\{
\begin{array}{ll}
\dot{u}=\{u,{\cal
H}\}=-\frac{1}{12}e^{-3u}p_u,\hspace{.5cm}\dot{p_u}=\{p_u,{\cal
H}\}=\frac{1}{8}e^{-3u}\left(-p_u^2+p_v^2+p_w^2\right)-3\Lambda e^{3u},\\\\
\dot{v}=\{v,{\cal
H}\}=\frac{1}{12}e^{-3u}p_v,\hspace{.5cm}\dot{p_v}=\{p_v,{\cal
H}\}=0,\\\\
\dot{w}=\{w,{\cal
H}\}=\frac{1}{12}e^{-3u}p_w,\hspace{.5cm}\dot{p_w}=\{p_w,{\cal
H}\}=0.
\end{array}
\right.
\end{eqnarray}We also have the constraint equation ${\cal H} = 0$, from which we obtain
\begin{equation}\label{P}
e^{-3u}\left(-p_u^2+p_v^2+p_w^2\right)=-24\Lambda
e^{3u}.\end{equation}Using this relation in eliminating $p_u$ from
the first equations of the system (\ref{O}) results
\begin{equation}\label{Q}
\ddot{u}+3\dot{u}^2=\frac{\omega^2}{3},\end{equation}where
$\omega=\sqrt{\frac{3\Lambda}{2}}$. The solution of this equation
can be read as
\begin{equation}\label{R}
u(t)=\frac{1}{3}C+\frac{1}{3}\ln \left[\sinh \omega
(t-\delta)\right],\end{equation}where $C$ and $\delta$ are
integrating constants. Since $\delta$ only corresponds to a shift
in the big-bang singularity, we can put it equal to zero. The last
two equations of the system (\ref{O}) indicate that $v(t)$ and
$w(t)$ obey the equations of motion
$\dot{v}=\frac{1}{12}p_{0v}e^{-3u}$ and
$\dot{w}=\frac{1}{12}p_{0w}e^{-3u}$, where we take
$p_v=p_{0v}=\mbox{cons}.$ and $p_w=p_{0w}=\mbox{cons}.$ Upon
substituting the relation (\ref{R}) into these equations, they can
easily be integrated to yield

\begin{equation}\label{S}
v(t)=\frac{p_{0v}}{12 e^C \omega}\ln \left[\tanh \frac{\omega
t}{2}\right], \hspace{0.5cm}w(t)=\frac{p_{0w}}{12 e^C \omega}\ln
\left[\tanh \frac{\omega t}{2}\right].\end{equation}Now, these
solutions must satisfy the zero energy condition, ${\cal H}=0$.
Thus, substitution of equations (\ref{R}) and (\ref{S}) into
(\ref{P}) gives a relation between integration constants as
\begin{equation}\label{S1}
C=\ln \frac{\sqrt{p_{0v}^2+p_{0w}^2}}{4\omega}.\end{equation} From
these relations one can calculate all of the physical quantities
of observational interest in cosmology. In analogy with a FRW
Universe, we can define a Hubble parameter corresponding to each
of the three different directions
\begin{equation}\label{C1}
H_a=\frac{\dot{a}}{a},\hspace{0.5cm}H_b=\frac{\dot{b}}{b},\hspace{0.5cm}H_c=\frac{\dot{c}}{c},\end{equation}in
terms of which the average Hubble parameter is defined as
\begin{equation}\label{C2}
H=\frac{1}{3}(H_a+H_b+H_c)=\dot{u}.\end{equation} The directional
Hubble parameter $H_i$ measures the expansion rate of the Universe
in the direction $x_i$ while the average Hubble parameter $H$
measures its volumetric expansion rate. We will also define the
expansion scalar $\Theta=3H$, the shear scalar
$\sigma^2=\frac{1}{2}\left(\sum_{i=1}^3
H_i^2-3H^2\right)=3\left(\dot{v}^2+\dot{w}^2\right)$, which
measures the degree of anisotropy of the space-time and the
average deceleration parameter $q=\frac{d}{dt}(\frac{1}{H})-1$,
where as is well known is indicated by how much the expansion of
the Universe is slowing down. If the expansion is speeding up, for
which there appears to be some recent evidence, then this
parameter will be negative. Now, let us return to the variables
$a_i(t)$ [$a_1(t)=a(t)$, $a_2(t)=b(t)$ and $a_3(t)=c(t)$] using
the transformation (\ref{A2}), in terms of which we obtain the
corresponding scale factors for the Bianchi I cosmology as
\begin{equation}\label{C3}
a_i(t)=\left(\frac{\sqrt{p_{0v}^2+p_{0w}^2}}{2\omega}\right)^{1/3}\sinh^{\frac{1}{3}+s_i}\frac{\omega
t}{2}\cosh^{\frac{1}{3}-s_i}\frac{\omega t}{2},\end{equation}where
the constants $s_i$ are defined as
\begin{equation}\label{C_4}
s_1=\frac{p_{0v}+\sqrt{3}p_{0w}}{3\sqrt{p_{0v}^2+p_{0w}^2}},\hspace{0.5cm}s_2=\frac{p_{0v}-\sqrt{3}p_{0w}}{3\sqrt{p_{0v}^2+p_{0w}^2}},\hspace{0.5cm}
s_3=\frac{-2p_{0v}}{3\sqrt{p_{0v}^2+p_{0w}^2}},\hspace{0.5cm}\end{equation}and
satisfy the following relations
\begin{equation}\label{C5}
s_1+s_2+s_3=0,\hspace{0.5cm}s_1^2+s_2^2+s_3^2=\frac{2}{3}.\end{equation}From
these equations we obtain, the directional Hubble parameters
\begin{equation}\label{C6}
H_i=\frac{\dot{a_i}}{a_i}=\frac{s_i \omega}{\sinh \omega
t}+\frac{\omega}{3}\coth \omega t,\end{equation}the comoving
volume
\begin{equation}\label{C7}
V(t)=\frac{\sqrt{p_{0v}^2+p_{0w}^2}}{4\omega}\sinh \omega
t,\end{equation}the average Hubble parameter
\begin{equation}\label{C8}
H(t)=\frac{\omega}{3}\coth \omega t,\end{equation}the shear scalar
\begin{equation}\label{C9}
\sigma(t)=\frac{\sqrt{3}}{4}\frac{1}{\sinh \omega
t},\end{equation}and the deceleration parameter
\begin{equation}\label{C10}
q(t)=\frac{3}{\cosh^2 \omega t}-1.\end{equation} In figure
\ref{fig1}, we have shown the behavior of the scale factors,
Hubble parameters, comoving volume, shear scalar and deceleration
parameter for typical values of the parameters. It is seen that
the evolution of the Universe begins with a big-bang singularity
at $t=0$ with a high degree of anisotropy and then follows an
expansion phase in all directions. Since we have negative
acceleration (positive deceleration parameter) at early times, the
Universe decelerates its volumetric expansion in this era. On the
other hand, the positive acceleration (negative deceleration
parameter) is occurred for late times which means that the
Universe currently accelerates its volumetric expansion. While the
volume of the Universe and its scale factors in different
directions increase monotonically, the shear scalar decreases
monotonically and tends to zero as $t\rightarrow \infty$. This
means at late time the model approaches that of the flat FRW
Universe with cosmological constant, i.e., the Universe eventually
evolves to a phase where it is close to a de Sitter solution.

\section{Lie-algebraically deformed phase space}
We now focus attention on the study of the effects of
Lie-algebraic noncommutativity on the cosmological dynamic of the
Bianchi type I model described above. This kind of
noncommutativity is described by a deformed Poisson algebra
between the phase space variables with the Lie-algebraic structure
as follows
\begin{equation}\label{T}
\{x_i,x_j\}=\theta^k_{ij}x_k,\end{equation}where the constants
$\theta^k_{ij}$ are assumed to be antisymmetric with respect to
their lower indices. To guarantee the Jacobi identity
\begin{equation}\label{T1}
\{\{x_i,x_j\},x_k\}+
\{\{x_j,x_k\},x_i\}+\{\{x_k,x_i\},x_j\}=0,\end{equation} these
parameters are also subject to the following conditions
\begin{equation}\label{U}
\theta^l_{ij}\theta^m_{lk}+\theta^l_{ki}\theta^m_{lj}+\theta^l_{jk}\theta^m_{li}=0.\end{equation}
Following \cite{Yan}, we consider the Poisson brackets between the
dynamical variables and their momenta as
\begin{equation}\label{V}
\{x_i,p_j\}=\delta_{ij}+\bar{\theta}^k_{ij}x_k+\tilde{\theta}^k_{ij}p_k,\hspace{0.5cm}\{p_i,p_j\}=0,\end{equation}in
which the new noncommutative parameters $\bar{\theta}$ and
$\tilde{\theta}$ with vanishing diagonal elements are introduced
and furthermore, we assume that the momenta (Poisson) commute with
each other. For our model at hand, $x_i(i=1,2,3)=v,w,u$ and
$p_i(i=1,2,3)=p_v,p_w,p_u$, as before. In the phase space with
coordinates $\vec{\eta}=({\bf x},{\bf p})$, in addition of the
identity (\ref{T1}), the following Jacobi identities should also
hold
\begin{equation}\label{V1}
\{\{x_i,x_j\},p_k\}+
\{\{x_j,p_k\},x_i\}+\{\{p_k,x_i\},x_j\}=0,\end{equation}

\begin{equation}\label{V2}
\{\{x_i,p_j\},p_k\}+
\{\{p_j,p_k\},x_i\}+\{\{p_k,x_i\},p_j\}=0.\end{equation}These
relations impose some more constraints on the noncommutative
parameters as
\begin{eqnarray}\label{V3}
\left\{
\begin{array}{ll}
\theta ^k_{ij}+\tilde{\theta}^j_{ik}-\tilde{\theta}^i_{jk}=0,\\\\
\theta^l_{ij}\bar{\theta}^m_{lk}+\bar{\theta}^l_{jk}\theta^m_{li}-\bar{\theta}^m_{il}\tilde{\theta}^l_{jk}-\bar{\theta}^l_{ik}\theta^m_{lj}
+\tilde{\theta}^l_{ik}\bar{\theta}^m_{jl}=0,\\\\
\theta^l_{ij}\tilde{\theta}^m_{lk}+\tilde{\theta}^l_{ik}\tilde{\theta}^m_{jl}-\tilde{\theta}^l_{jk}\tilde{\theta}^m_{il}=0,\\\\
\bar{\theta}^k_{ij}-\bar{\theta}^j_{ik}=0,\\\\
\bar{\theta}^l_{ij}\bar{\theta}^m_{lk}-\bar{\theta}^l_{ik}\bar{\theta}^m_{lj}=0,\\\\
\bar{\theta}^l_{ij}\tilde{\theta}^m_{lk}-\bar{\theta}^l_{ik}\tilde{\theta}^m_{lj}=0.
\end{array}
\right.
\end{eqnarray}
Now, the equations (\ref{U}) and (\ref{V3}) should be solved to
determine the noncommutative parameters. Here we will not deal
with the details of the methods through which one can find these
solutions and only refer to the Refs. \cite{Yan} where the
following two kinds of solutions to the equations (\ref{U}) and
(\ref{V3}) are proposed.

$\bullet$ Type I: For this type of solution, the nonvanishing
noncommutative parameters are as follows (the antisymmetric
counterparts of $\theta^k_{ij}$ should be also considered)
\begin{equation}\label{W}
\theta^2_{13}=-\theta^1_{23}=\theta,\hspace{0.5cm}\tilde{\theta}^1_{32}
=-\tilde{\theta}^2_{31}=\theta.\end{equation} In this case the
Poisson brackets of the phase space variables correspond to our
cosmological setting read as
\begin{eqnarray}\label{AB}
\left\{
\begin{array}{ll}
\{v,w\}=0,\hspace{0.5cm}\{w,u\}=-\theta v,\hspace{0.5cm}\{u,v\}=-\theta w,\\\\
\{u,p_u\}=1,\hspace{0.5cm}\{u,p_v\}=-\theta p_w,\hspace{0.5cm}\{u,p_w\}=\theta p_v,\\\\
\{v,p_u\}=0,\hspace{0.5cm}\{v,p_v\}=1,\hspace{0.5cm}\{v,p_w\}=0,\\\\\textbf{
}\{w,p_u\}=0,\hspace{0.5cm}\{w,p_v\}=0,\hspace{0.5cm}\{w,p_w\}=1
\end{array}
\right.
\end{eqnarray}

$\bullet$ Type II: For this type of solution, we have the
following nonvanishing noncommutative parameters (and also the
antisymmetric counterparts of $\theta^{k}_{ij}$)
\begin{equation}\label{AC}
\theta^2_{13}=-\theta^1_{23}=\theta,\hspace{0.5cm}
\tilde{\theta}^1_{32}=-\tilde{\theta}^2_{31}=\theta,\hspace{0.5cm}
\bar{\theta}^2_{31}=\bar{\theta}^1_{32}=-\bar{\theta}.\end{equation}
In this case the Poisson brackets of the phase space variables
correspond to our cosmological setting read as
\begin{eqnarray}\label{AD}
\left\{
\begin{array}{ll}
\{v,w\}=0,\hspace{0.5cm}\{w,u\}=-\theta v,\hspace{0.5cm}\{u,v\}=-\theta w,\\\\
\{u,p_u\}=1,\hspace{0.5cm}\{u,p_v\}=-\bar{\theta}w-\theta p_w,\hspace{0.5cm}\{u,p_w\}=-\bar{\theta} v+\theta p_v,\\\\
\{v,p_u\}=0,\hspace{0.5cm}\{v,p_v\}=1,\hspace{0.5cm}\{v,p_w\}=0,\\\\
\{w,p_u\}=0,\hspace{0.5cm}\{w,p_v\}=0,\hspace{0.5cm}\{w,p_w\}=1
\end{array}
\right.
\end{eqnarray}The Poisson bracket for any phase space function can be straightforward and obtained
from (\ref{AB}) and (\ref{AD}) and reads
\begin{equation}\label{AE}
\{f({\bf x},{\bf p}),g({\bf x},{\bf
p})\}=\{\eta^A,\eta^B\}\frac{\partial f}{\partial
\eta^A}\frac{\partial g}{\partial \eta^B},\end{equation}where
$\vec{\eta}=({\bf x},{\bf p})=(v,w,u,p_v,p_w,p_u)$. The time
evolution of such a function is thus given by
\begin{equation}\label{AF}
\frac{df}{dt}=\{f,{\cal H}\}.\end{equation}

\begin{figure}
\begin{tabular}{c}\hspace{-1cm} \epsfig{figure=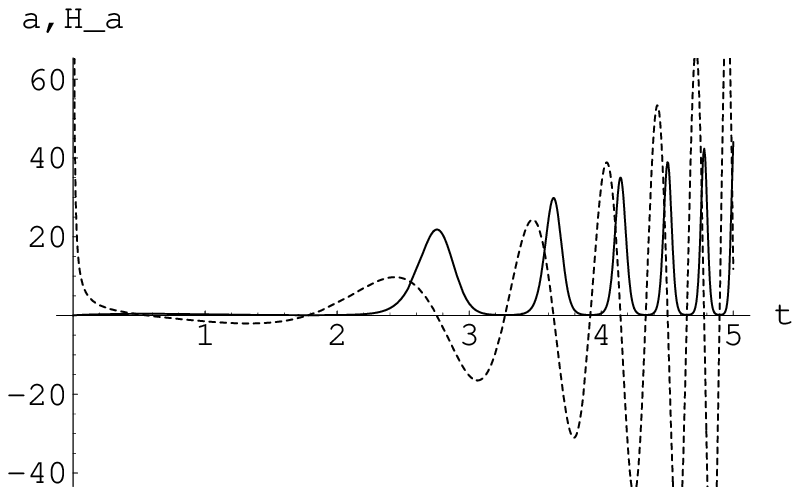,width=5cm}
\hspace{1.5cm} \epsfig{figure=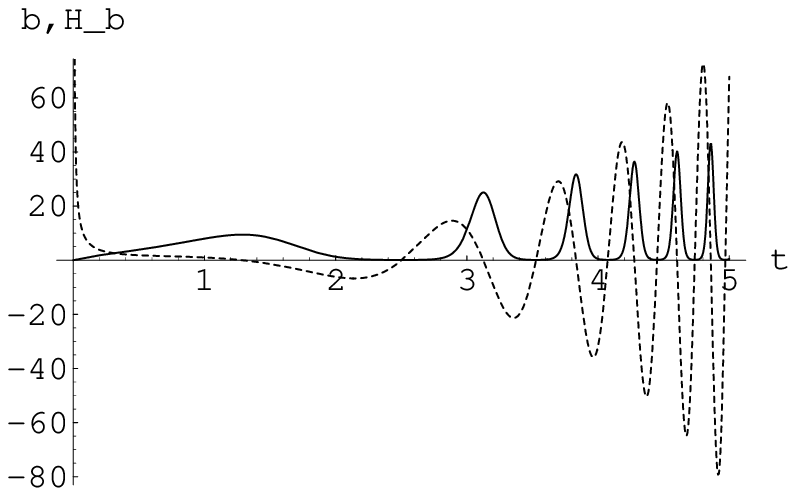,width=5cm}\hspace{1.5cm}
\epsfig{figure=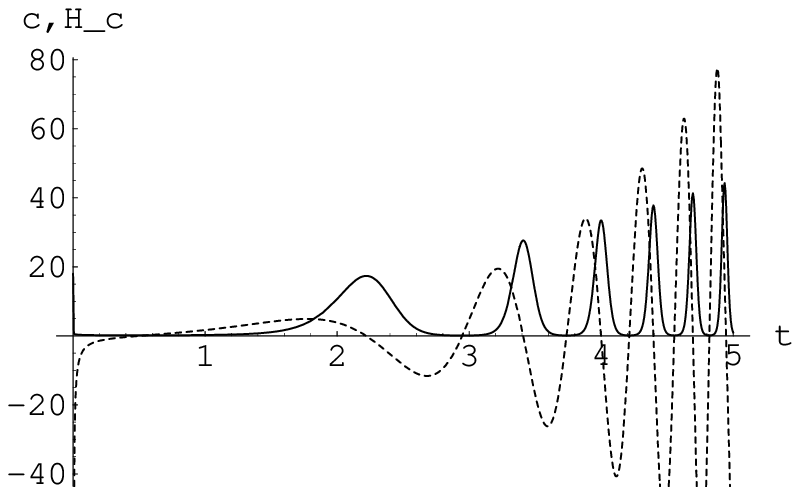,width=5cm}
\end{tabular}
\begin{tabular}{c}\hspace{-1cm} \epsfig{figure=fig4.eps,width=5cm}
\hspace{1.5cm} \epsfig{figure=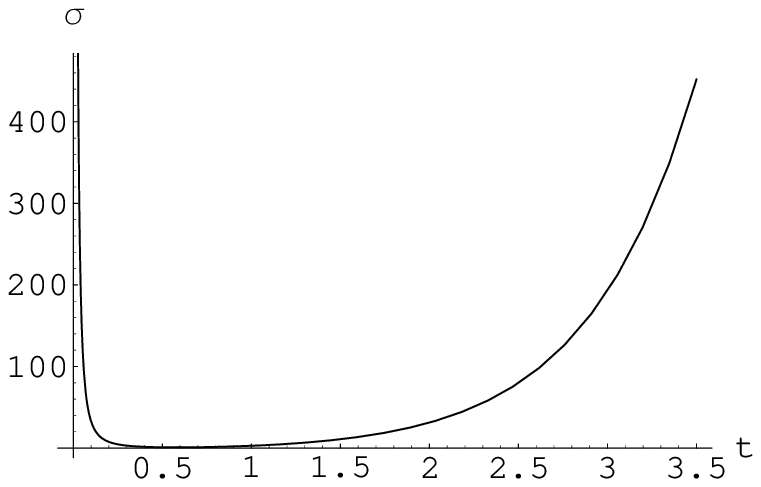,width=5cm}\hspace{1.5cm}
\epsfig{figure=fig6.eps,width=5cm}
\end{tabular}
\caption{\footnotesize Top: Behavior of the scale factors $a(t)$,
$b(t)$ and $c(t)$ (solid lines) and their corresponding Hubble
parameters (dashed lines) as a function of time. Bottom: The
comoving volume $V(t)=e^{3u}=abc$ (solid line) and its
corresponding Hubble parameter (dashed line), the shear scalar and
the deceleration parameter. The figures are plotted for numerical
values $\Lambda=1$, $p_{0v}=1$, $p_{0w}=2$, $\theta=0.25$ and
$C_1=C_2=1$.}\label{fig2}
\end{figure}

For deformed phase space of type I the cosmological dynamics of
the Bianchi-I model are
\begin{eqnarray}\label{AG}
\left\{
\begin{array}{ll}
\dot{u}=\{u,{\cal H}\}=-\frac{1}{12}e^{-3u}p_u,\\\\
\dot{v}=\{v,{\cal H}\}=\theta w \left[-\frac{1}{8}e^{-3u}\left(-p_u^2+p_v^2+p_w^2\right)+3\Lambda e^{3u}\right]
+\frac{1}{12}e^{-3u}p_v,\\\\
\dot{w}=\{w,{\cal H}\}=-\theta v\left[-\frac{1}{8}e^{-3u}\left(-p_u^2+p_v^2+p_w^2\right)+3\Lambda e^{3u}\right]
+\frac{1}{12}e^{-3u}p_w,\\\\
\dot{p_u}=\{p_u,{\cal H}\}=-3\Lambda e^{3u}+\frac{1}{8}e^{-3u}\left(-p_u^2+p_v^2+p_w^2\right),\\\\
\dot{p_v}=\{p_v,{\cal H}\}=\theta p_w \left[-\frac{1}{8}e^{-3u}\left(-p_u^2+p_v^2+p_w^2\right)+3\Lambda e^{3u}\right],\\\\
\dot{p_w}=\{p_w,{\cal H}\}=-\theta
p_v\left[-\frac{1}{8}e^{-3u}\left(-p_u^2+p_v^2+p_w^2\right)+3\Lambda
e^{3u}\right].
\end{array}
\right.
\end{eqnarray}As before, assuming the full Einstein field equations hold,
this implies that the corresponding Hamiltonian must vanish, that
is ${\cal H}=0$, which yields
\begin{equation}\label{AH}
\frac{1}{8}e^{-3u}\left(-p_u^2+p_v^2+p_w^2\right)=-3\Lambda
e^{3u}.\end{equation}We see that the noncommutativity does not
affect the dynamics of $u$ and $p_u$. Therefore, we have
\begin{equation}\label {AI}
u(t)=\frac{1}{3}C+\frac{1}{3}\ln[\sinh \omega
t],\hspace{0.5cm}p_u(t)=-4\omega e^C \cosh \omega t,\end{equation}
which means that the volumetric expansion of the Universe given by
$V(t)=e^{3u(t)}$ is the same as the ordinary Bianchi I model
described in the previous section. Substituting these results into
the rest of the equations of the system (\ref{AG}), we are led to
\begin{eqnarray}\label{AJ}
\left\{
\begin{array}{ll}
\dot{v}=6\Lambda \theta e^C w \sinh \omega t+\frac{1}{12 e^C \sinh \omega t}p_v,\\\\
\dot{w}=-6\Lambda \theta e^C v \sinh \omega t +\frac{1}{12 e^C \sinh \omega t}p_w,\\\\
\dot{p_v}=6\Lambda \theta e^C p_w \sinh \omega t,\\\\
\dot{p_w}=-6\Lambda \theta e^C p_v \sinh \omega t.
\end{array}
\right.
\end{eqnarray}From the third and fourth equations of this system, we obtain
\begin{equation}\label{AL}
\ddot{p_v}-\omega \dot{p_v}\coth \omega t +36\Lambda^2\theta^2
e^{2C}p_v \sinh^2 \omega t=0,\end{equation} where after
integration we get the following expressions for $p_v$ and $p_w$
\begin{equation}\label{AM}
p_v(t)=p_{0v}\cos \left(\frac{6\Lambda \theta e^C}{\omega}\cosh
\omega t\right)+p_{0w}\sin \left(\frac{6\Lambda \theta
e^C}{\omega}\cosh \omega t\right),\end{equation}
\begin{equation}\label{AN}
p_w(t)=-p_{0v}\sin \left(\frac{6\Lambda \theta e^C}{\omega}\cosh
\omega t\right)+p_{0w}\cos \left(\frac{6\Lambda \theta
e^C}{\omega}\cosh \omega t\right),\end{equation}in which we have
set the integration constants as $p_{0v}$ and $p_{0w}$ to achieve
the correct results in the limit $\theta \rightarrow 0$. With
these results at hand we can integrate the two first equations of
the system (\ref{AJ}) to obtain

\begin{eqnarray}\label{AO}
v(t)&=&C_1\cos \left(\frac{6\Lambda \theta e^C }{\omega}\cosh
\omega t\right)+C_2\sin \left(\frac{6\Lambda \theta e^C
}{\omega}\cosh \omega t\right)+\nonumber \\ && \frac{1}{12 e^C
\omega}\ln \left(\tanh \frac{\omega t}{2}\right)\left\{p_{0v}\cos
\left(\frac{6\Lambda \theta e^C }{\omega}\cosh \omega
t\right)+p_{0w}\sin \left(\frac{6\Lambda \theta e^C }{\omega}\cosh
\omega t\right)\right\},
\end{eqnarray}

\begin{eqnarray}\label{AP}
w(t)&=&C_2\cos \left(\frac{6\Lambda \theta e^C }{\omega}\cosh
\omega t\right)-C_1\sin \left(\frac{6\Lambda \theta e^C
}{\omega}\cosh \omega t\right)+\nonumber \\ &&\frac{1}{12 e^C
\omega}\ln \left(\tanh \frac{\omega t}{2}\right)\left\{p_{0w}\cos
\left(\frac{6\Lambda \theta e^C }{\omega}\cosh \omega
t\right)-p_{0v}\sin \left(\frac{6\Lambda \theta e^C }{\omega}\cosh
\omega t\right)\right\},
\end{eqnarray}where $C_1$ and $C_2$ are integration constants. The requirement that the deformed Hamiltonian constraints (\ref{AH})
should hold during the evolution of the system leads again to the
relation (\ref{S1}) between integrating constants. In figure
\ref{fig2} we have plotted the cosmological functions
corresponding to the above solutions. As is clear from this figure
the volumetric expansion, average Hubble and deceleration
parameters of the deformed model are exactly the same as the usual
Bianchi I model. This is because these parameters come from the
function $u(t)$ which is not affected by the deformation on the
phase space. However, while like the usual model, the volume of
the Universe increases continuously, the evolution of the scale
factors in three directions is quite different in comparison with
the usual Bianchi I model. We see that the scale factors behave
cyclically, i.e., they  undergo a periodic sequence of
contractions and expansions phases. In contrast to the
conventional big-bang inflationary cosmology described in the
previous section in which the Universe begins with the big-bang
singularity and expands forever, here we have an alternative in
which the bang is replaced by a transition to an earlier phase of
evolution. In each cycle of such a cosmology the scale factors
begin their evolution with  a big-bang and end in a big-crunch,
only to emerge in a big-bang once again. Cyclic cosmologies have
been extensively studied in literature in an attempt to solve some
problems of the standard cosmological model such as the flatness
problem, the horizon problem, the coincidence problem, the problem
of initial conditions etc. \cite{10}. In general, a cyclic
Universe can be generated by adding a matter or a field that will
produce a bounce, and then explore what conditions are to be
imposed on it to produce oscillations. In the above analysis we
provided an alternative method to construct a Bianchi type I
cosmology with cyclic scale factors by introducing a deformation
of Lie-algebraic type on the phase space of this model. In our
model although the expansion of the whole volume of the Universe
is speeding up, for which there appears to be some recent
evidence, from the evolution of the scale factors point of view,
the history of the Universe is periodic and all of the key events
that construct the large scale structure of the observable
Universe occurred a cycle ago. An important feature of this model
is that its shear scalar which measures the degree of anisotropy
in the space-time, behaves nonmonotonically. For small values of
$t$, it decreases and tends to zero and then it increases
monotonically, eventually diverging as $t\rightarrow \infty$.
Therefore, the model under consideration does not satisfy the
conditions for approaching isotropy. Since this behavior is not
consistent with late time observations, it seems an additional
mechanism is needed to modify this feature of the model. This is
the subject of our forthcoming work and thus we do not deal with
this issue here.

\begin{figure}
\begin{tabular}{c}\hspace{-1cm} \epsfig{figure=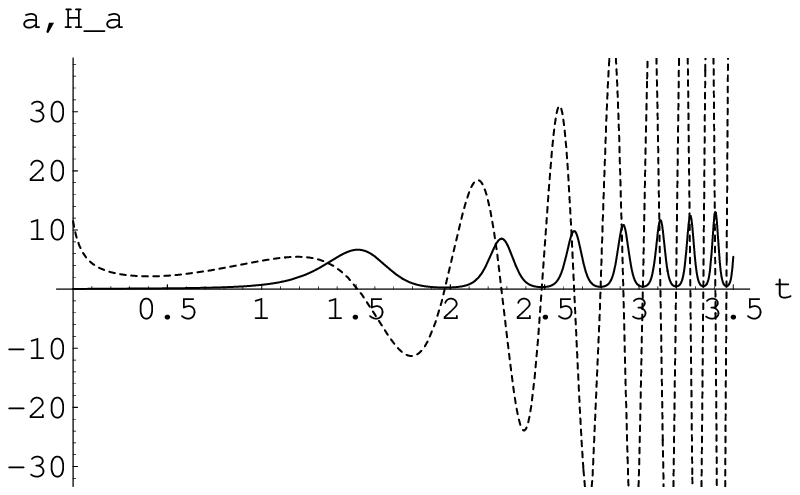,width=5cm}
\hspace{1.5cm} \epsfig{figure=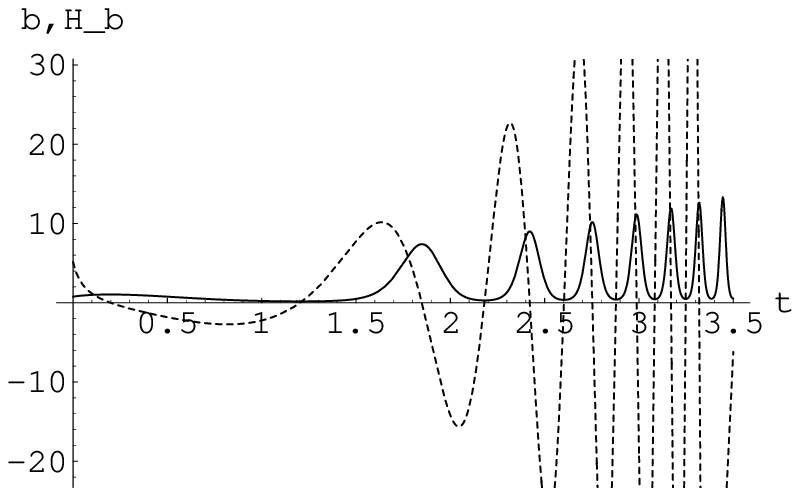,width=5cm}\hspace{1.5cm}
\epsfig{figure=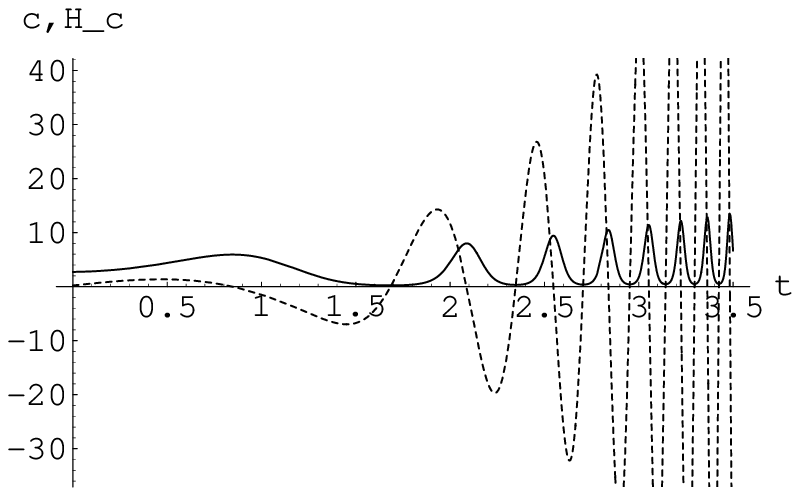,width=5cm}
\end{tabular}
\begin{tabular}{c}\hspace{-1cm} \epsfig{figure=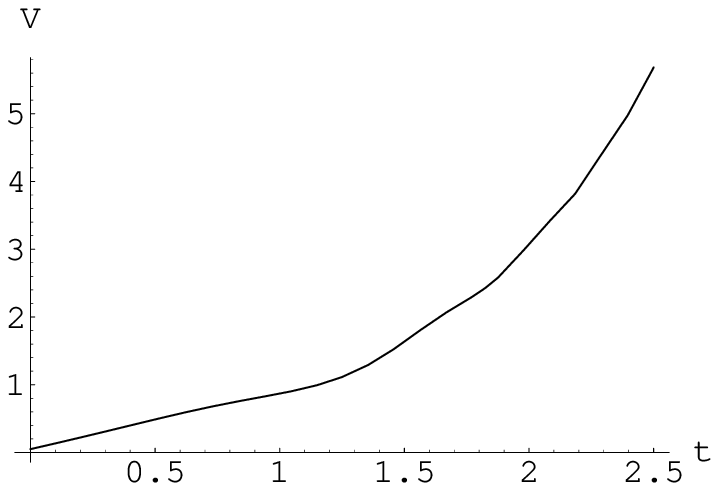,width=5cm}
\hspace{1.5cm} \epsfig{figure=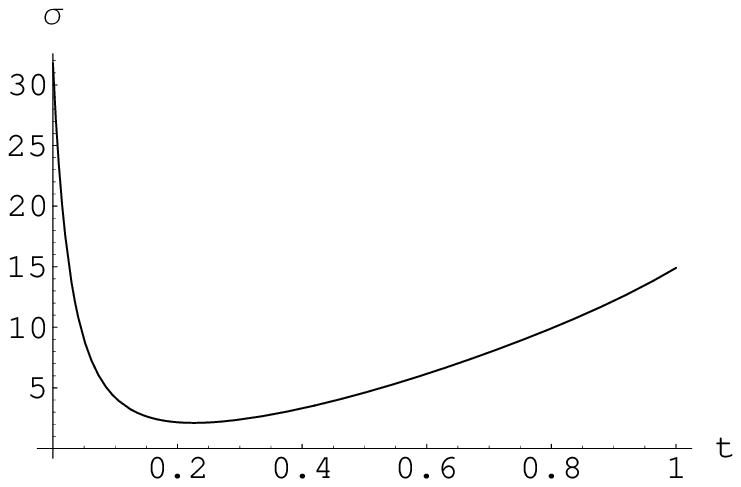,width=5cm}\hspace{1.5cm}
\epsfig{figure=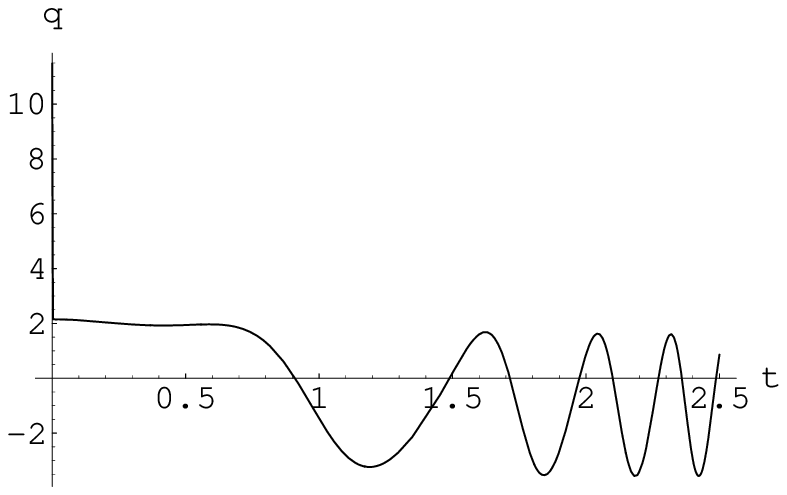,width=5cm}
\end{tabular}
\caption{\footnotesize Up: Qualitative behavior of the scale
factors $a(t)$, $b(t)$ and $c(t)$ (solid lines) and their
corresponding Hubble parameters (dashed lines) as a function of
time. Bottom: The comoving volume $V(t)=e^{3u}=abc$ , the shear
scalar and the deceleration parameter. The figures are plotted for
numerical values $\Lambda=1$, $\theta=0.5$ and $\bar{\theta}=0.5$.
We take the initial conditions $u(0)=v(0)=w(0)=-1$, $p_{0u}=-2$,
$p_{0v}=1.7$ and $p_{0w}=1$. After examining other sets of initial
conditions, we verify that this behavior repeats
itself.}\label{fig3}
\end{figure}

For deformed phase space of type II, using the (Poisson)
commutation relations (\ref{AD}), the cosmological dynamics of the
Bianchi-I model after applying the Hamiltonian constraint are
\begin{eqnarray}\label{AQ}
\left\{
\begin{array}{ll}
\dot{u}=\{u,{\cal H}\}=-\frac{1}{12}\bar{\theta}e^{-3u}(wp_v+vp_w)-\frac{1}{12}e^{-3u}p_u,\\\\
\dot{v}=\{v,{\cal H}\}=6\Lambda \theta w e^{3u}+\frac{1}{12}e^{-3u}p_v,\\\\
\dot{w}=\{w,{\cal H}\}=-6\Lambda \theta v e^{3u}+\frac{1}{12}e^{-3u}p_w,\\\\
\dot{p_u}=\{p_u,{\cal H}\}=-6\Lambda e^{3u},\\\\
\dot{p_v}=\{p_v,{\cal H}\}=6\Lambda (\bar{\theta}w+\theta p_w)e^{3u},\\\\
\dot{p_w}=6\Lambda (\bar{\theta}v-\theta p_v)e^{3u}.
\end{array}
\right.
\end{eqnarray}We see that this kind of deformed cosmology forms a system of nonlinear coupled differential
equations which unfortunately cannot be solved analytically. In
figure \ref{fig3}, employing numerical methods, we have shown the
approximate behavior of the corresponding cosmological functions
for typical values of the parameters and initial conditions
respectively. From this figure it is seen that the cyclic behavior
of the corresponding cosmology is almost like the case of type I
deformation except that here the average deceleration parameter
also behaves oscillatory which means that the Universe experiences
a periodic sequence in which it decelerates and accelerates its
expansion alternatively.

\section{Conclusions}
In this paper we have studied the possibility of emergence of a
Bianchi type I cyclic cosmology in the framework of a deformed
phase space of Lie-algebraic type. In the proposed model the phase
space coordinates obey the (Poisson) commutation relations with a
Lie-algebraic structure, the structure constants of which play the
role of the deformation parameters. These constants are subject to
conditions to guarantee the Jacobi identity which by solving them
the noncommutative (deformation) parameters can be found. In this
line we have followed Refs. \cite{Yan} that offer two types of
noncommutative parameters which satisfy the corresponding
constraints. We investigated the Bianchi type I cosmology based on
these two kinds of deformed phase space and compare the results
with the conventional model. For the deformed model of the first
kind we showed that while the behavior of the whole volume of the
Universe and its  deceleration parameter are the same as the
consensus Bianchi I model, the evolution of its scale factors is
quite different. In spite of the usual Bianchi I model in which
the three scale factors increase monotonically, in the
Lie-algebraically deformed model they evolve cyclically. This
means that in the scenario proposed by the deformed model the
scale factors of the Universe undergo an endless sequence of
epochs which begin with a big-bang and end in a big-crunch, i.e.,
the size of the Universe in each direction bounces from
contraction to reexpansion alternatively. We saw that this
behavior causes a high degree of anisotropy as $t\rightarrow
\infty$ in contrast to the late time observations and therefore it
seems an additional mechanism is needed to modify this feature of
the model. For the second kind of deformation parameters, since we
could not solve the system of field equations analytically, we
investigated the subject by employing numerical methods. Our
numerical analysis in this case showed again an oscillatory
behavior not only for the scale factors but also for the total
deceleration parameter. \vspace{5mm}\newline \noindent {\bf
Acknowledgements}\vspace{2mm}\noindent\newline The authors are
grateful to the anonymous referees for enlightening suggestions.
B. V. would like to thank the research council of Azad University
of Chalous for financial support.

\end{document}